\documentclass[prd,aps]{revtex4-2}
\usepackage{amsmath}
\usepackage{amssymb}
\usepackage{amsfonts}
\usepackage{graphicx,bm}
\usepackage{dcolumn}
\usepackage[colorlinks=true]{hyperref}
\usepackage{epsf}
\usepackage{enumerate}
\usepackage{hhline}
\usepackage{array}
\usepackage{tabularx}
\usepackage{subfigure}

\newcommand{\be}{\begin{equation}}
\newcommand{\ee}{\end{equation}}
\newcommand{\bea}{\begin{eqnarray}}
\newcommand{\eea}{\end{eqnarray}}
\newcommand{\beaa}{\begin{eqnarray*}}
\newcommand{\eeaa}{\end{eqnarray*}}

\newcommand{\nn}{\nonumber \\}
\newcommand{\e}{\mathrm{e}}

\newcommand{\dd}{{\rm d}}

\def\be{\begin{equation}}
\def\ee{\end{equation}}
\def\bea{\begin{eqnarray}}
\def\eea{\end{eqnarray}}

\begin{document}
\title{Eternal vs singular observers in interacting dark energy-dark matter models}
\author{Diego \'Alvarez-Ortega} \email{diego.alvarezo@alumnos.unican.es}
\affiliation{Instituto de Física de Cantabria (CSIC - UC), Avda. de los Castros s/n,
39005 - Santander, Spain}
	\author{Gonzalo J. Olmo} \email{gonzalo.olmo@uv.es}
\affiliation{Departamento de F\'{i}sica Te\'{o}rica and IFIC, Centro Mixto Universidad de Valencia - CSIC.
Universidad de Valencia, Burjassot-46100, Valencia, Spain}
\affiliation{Departamento de F\'isica, Universidade Federal do Cear\'a (UFC), Campus do Pici, C.P. 6030, Fortaleza, 60455-760, Brazil}
\author{Diego Rubiera-Garcia} \email{drubiera@ucm.es}
\affiliation{Departamento de F\'isica Te\'orica and IPARCOS,
	Universidad Complutense de Madrid, E-28040 Madrid, Spain}
\author{Diego S\'aez-Chill\'on G\'omez}
\email{diego.saez@uva.es} \affiliation{Department of Theoretical Physics, Atomic and Optics, Campus Miguel Delibes, \\ University of Valladolid UVA, Paseo Bel\'en, 7,
47011 - Valladolid, Spain}

\begin{abstract}
Interacting dark energy-dark matter models have been widely analyzed in the literature in an attempt to find traces of new physics beyond the usual cosmological ($\Lambda$CDM) models. Such a coupling between both dark components is usually introduced in a phenomenological way through a flux in the continuity equation. However, models with a  Lagrangian formulation are also possible. A class of the latter assumes a conformal/disformal coupling that leads to a fifth force on the dark matter component, which consequently does not follow the same geodesics as the other (baryonic, radiation, and dark energy) matter sources. Here we analyze how the usual cosmological singularities of the standard matter frame are seen from the dark matter one, concluding that by choosing an appropriate coupling,  dark matter observers will see no singularities but a non-beginning, non-ending universe. By considering two simple phenomenological models we show that such a type of coupling can fit observational data as well as the usual $\Lambda$CDM model.
\end{abstract}

\maketitle
%
%
%
\section{Introduction}
\label{intro}

The understanding of space-time singularities continues to be one of the major challenges in theoretical gravitational physics. The theorems developed by Penrose and Hawking (among others) over the seventies provided a formal and clear criterion to determine the occurrence of a spacetime singularity by making use of the notion of geodesic incompleteness \cite{Theorems}. Working under very broad (and reasonable) assumptions on the structure of space-time, these theorems prove that singularities in the fabric of space-time are ubiquitous in General Relativity (for a pedagogical discussion see \cite{Curiel}), from the ones hidden behind the event horizon of black holes to the initial cosmological (Big Bang) singularity of the concordance ($\Lambda$CDM) model \cite{BigBang}. After the finding of the cosmological late-time acceleration \cite{Acc}, other singularities were added to this pool,  which may occur under specific models of dark energy, particularly when the latter violates some of the energy conditions \cite{Nojiri:2005sx,TV,BeltranJimenez:2016dfc}. Among these  new singularities, the most worrisome one is the so-called Big Rip \cite{BigRip}, in which the divergence of the scale factor drives all geodesics to end at at future finite cosmic time as measured by a comoving observer in a FLRW universe.  Other future singularities are not geodesically incomplete but can still drive some other magnitudes (such as the the Hubble factor, their derivatives, or the energy density and the pressure of the matter fields) to diverge at a future time. This has led the community to establish a general classification of such singularities according to the degree of  harm (or lack of it) inflicted upon every binding structure that can be caused for instance, by the occurrence of arbitrarily large tidal forces \cite{Criteria}. 

It has been shown that the development of a future singularity might be in agreement with the observational data, since the corresponding fits can be statistically as good as more standard cosmological models \cite{BeltranJimenez:2016fuy}. Nonetheless, as opposed to the ordinary Big Bang singularity that occurs under very broad conditions and is intrinsic to any space-time with an ordinary content of matter, the Big Rip singularity requires the violation of the Null Energy Condition, $\rho+p>0$, the so-called phantom fluid. The realization of such equation of state can be easily achieved in several ways, through non-canonical scalar fields \cite{Elizalde:2004mq,Nojiri:2005pu,Elizalde:2008yf}, modified gravities \cite{Nojiri:2009kx} or bulk viscosity \cite{Nojiri:2005sr}, among others. In addition, this singularity might arise in models of interacting dark energy-dark matter, where the coupling between both dark components is the responsible in inducing the singularity \cite{BeltranJimenez:2016dfc}.

Interacting dark energy-dark matter models have been widely analyzed in the literature on the grounds of exploring new features to discriminate among the many existing cosmological models (for a review see \cite{Wang:2016lxa}). The usual way  to introduce such a coupling between dark components relies on more phenomenological assumptions than theoretical ones, just by considering a varying function usually dubbed as $Q$, that characterizes the flux of energy between both dark components \cite{Zimdahl:2001ar}. As a consequence, the conservation of the energy-momentum tensor does not hold separately for each component but for the pair together. A different approach is to introduce the coupling at the level of the action and then work out the corresponding field equations \cite{Koivisto:2005nr,Amendola:1999qq,Amendola:2003wa,Amendola:2003eq}. In the simplest case the interaction is mediated by a function of the scalar field (dark energy) in front of the dark matter Lagrangian \cite{Koivisto:2005nr}, while in non-minimally coupled scalar field models the interaction naturally arises under a conformal transformation \cite{Amendola:1999qq,Amendola:2003wa}. As a generalization of the latter, some models assume a dark matter Lagrangian coupled to a metric tensor that is related to the space-time metric by conformal and disformal transformations that depend on the dark energy (scalar) field \cite{Zumalacarregui:2010wj,Sakstein:2014isa,vandeBruck:2013yxa,vandeBruck:2015rma,vandeBruck:2016jgg,vandeBruck:2016hpz,Dusoye:2020wom}. Such a type of models have been widely analyzed in the literature as they can provide a wide range of parametrizations of different dark energy models \cite{Zumalacarregui:2010wj} while being able to satisfy local and cosmological constraints \cite{Sakstein:2014isa,vandeBruck:2013yxa,vandeBruck:2016hpz}.  Despite the fact that disformal couplings were originally introduced for dark matter, these can be easily extended to multiple fluids \cite{vandeBruck:2016jgg}, and may even induce changes in the fine-structure constant when the electromagnetic field is incorporated to the model \cite{vandeBruck:2015rma}. Moreover, such models can be constructed in such a way to mimic the $\Lambda$CDM model so as to satisfy current observational data \cite{Dusoye:2020wom}.

The present paper is devoted to the occurrence of future singularities in interacting dark energy-dark matter models involving conformal and disformal transformations. The fact that in such models the dark matter Lagrangian is coupled to a metric tensor that is different from the metric seen by ordinary matter sources (and dark energy alike), implies that that the former does not follow geodesics of the latter, being instead affected by a sort of fifth force that is the result of the interaction between the two dark components. However, since such a force just affects the dark matter component, this leaves our observational constraints untouched, imposing instead some limitations on the free parameters of the model due to the dynamics of the cosmological expansion, particularly on the scalar potential as well as on the conformal/disformal couplings, similarly to any cosmological model \cite{vandeBruck:2016hpz}. These features raise immediately the question on the issue of geodesic completeness in these models, i.e., when a space-time singularity occurs (and in which frame) and what does it entail for their observational viability. The main aim of the present paper is therefore to consider the two most worrying types of cosmological singularities (the Big Bang and Big Rip, respectively) as seen from the frame of the ordinary observer  (which therefore suffer from geodesic incompleteness), and to show that by a suitable conformal transformation such singularities can go away in the dark matter frame, i.e., the latter observers see a perfectly regular space-time. We dub this duality of viewpoints as singular versus eternal observers. Furthermore, by considering two simple phenomenological models we show that such idea can be promoted as a viable cosmological model by providing a fitting of its parameters according to Supernovae IA and BAO data that render them as compatible with current data as the $\Lambda$CDM model. 

The paper is organized as follows: in Sec. \ref{background} we review the formalism of conformal and disformal couplings in interacting dark energy-dark matter models. Sec. \ref{Recons} is devoted to establish a procedure to construct cosmological solutions starting from a particular theoretical model. In Sec. \ref{singu}, the analysis that leads to singular and regular expansions depending on the observer is performed and some analytical models are reconstructed. Then, in Sec. \ref{Phenom} we consider two simple phenomenological models that can be reproduced by the interacting models, which are subsequently compared with observational data in Sec. \ref{results}. Finally, Sec. \ref{conclusions} gathers the conclusions of the paper.

\section{Conformal and disformal couplings}
\label{background}

Let us consider the following gravitational action sourced by the corresponding fields contained in our hypothetical universe:
\be
S=\int d^4x \sqrt{-g} \left[\frac{R}{2\kappa^2}-\frac{1}{2}\partial_{\mu}\phi\partial^{\mu}\phi- V(\phi)+\mathcal{L}_{SM}(g,\psi_m)\right]+\int d^4x\sqrt{-\tilde{g}}\tilde{\mathcal{L}}_{DM}(\tilde{g},\tilde{\psi}_m)\ ,
\label{action}
\ee
where $\kappa^2=8\pi G$ is Newton's constant in suitable units, $\mathcal{L}_{SM}$ refers to the Lagrangian for Standard Model (SM) fields $\psi_m$ coupled to the metric $g_{\mu\nu}$, whereas $\tilde{\mathcal{L}}_{DM}$ is the Lagrangian for the dark matter fields $\tilde{\psi}_m$ coupled to another metric $\tilde{g}_{\mu\nu}$. The scalar field $\phi$ with its kinetic and potential terms will play the role of dark energy. Since in these action SM particles  and dark matter are coupled to different metrics, they will also follow different sets of geodesics, in such a way that the cosmological evolution might be (and will be) different as observed from each frame. Interacting dark energy-dark matter models based on the action (\ref{action}) assume that both metrics are related by a disformal transformation, which introduces a coupling term among the scalar field and dark matter as \cite{Zumalacarregui:2010wj,Sakstein:2014isa,vandeBruck:2013yxa,vandeBruck:2015rma,vandeBruck:2016jgg,vandeBruck:2016hpz,Dusoye:2020wom}:
\be
\tilde{g}_{\mu\nu}=C(\phi)g_{\mu\nu}+D(\phi)\partial_{\mu}\phi\partial_{\nu}\phi\ ,
\label{disformal}
\ee
where the functions $C(\phi)$ and $D(\phi)$ realize the conformal and disformal parts of the transformation, respectively. 

The field equations for the action (\ref{action}) are obtained by varying the action with respect to the metric $g_{\mu\nu}$, leading to:
\be
R_{\mu\nu}-\frac{1}{2}g_{\mu\nu}R=\kappa^2\left(T_{\mu\nu}^{SM}+T_{\mu\nu}^{\phi}+T_{\mu\nu}^{DM}\right)\ ,
\label{Fieldeqs}
\ee
where the corresponding energy-momentum tensors for each set of matter fields are defined as follows:
\bea
T_{\mu\nu}^{SM}&=&\frac{-2}{\sqrt{-g}}\frac{\delta \left(\sqrt{-g}\mathcal{L}_{SM}\right)}{\delta g^{\mu\nu}}\ , \nn
T_{\mu\nu}^{\phi}&=&\partial_{\mu}\phi\partial_{\nu}\phi-g_{\mu\nu}\left(\frac{1}{2}\partial_{\sigma}\phi\partial^{\sigma}\phi+V(\phi)\right)\ , \nn
T_{\mu\nu}^{DM}&=&\frac{-2}{\sqrt{-g}}\frac{\delta \left(\sqrt{-\tilde{g}}\tilde{\mathcal{L}}_{SM}\right)}{\delta g^{\mu\nu}}\ , 
\label{EMtensors}
\eea
for the SM particles, dark energy, and dark matter, respectively. On the other hand, variation with respect to the scalar field $\phi$ provides the equation:
\be
\Box\phi-\frac{dV(\phi)}{d\phi}=-Q\ ,
\label{ScalarFeq}
\ee
where  $Q$ accounts for the interacting term, given by
\be
Q=\frac{C'(\phi)}{2C(\phi)}g^{\mu\nu}T^{DM}_{\mu\nu}+\frac{D'(\phi)}{2C(\phi)}T^{DM\mu\nu}\partial_{\mu}\phi\partial_{\nu}\phi-\nabla_{\mu}\left[\frac{D(\phi)}{C(\phi)}T^{DM\mu\nu}\nabla_{\nu}\phi\right]\ .
\label{Qterm}
\ee
By computing the divergence of the energy-momentum tensor for SM and dark matter particles
\be
\nabla^{\mu}T_{\mu\nu}^{SM}=0\ , \quad \nabla^{\mu}T_{\mu\nu}^{DM}=Q\nabla_{\nu}\phi\ ,
\label{ContEqs}
\ee
one finds the energy conservation of the SM particles and the statement on the transfer of energy between the dark matter and the (dark energy) scalar field depending on the $Q$-term defined in Eq.(\ref{Qterm}). Combined with the scalar field equation (\ref{ScalarFeq}), these equations ensure the divergence of the right-hand side of the field equations (\ref{Fieldeqs}) to be null and, consequently, the energy conservation of the full matter sources (SM+dark energy+dark matter).

For the sake of the cosmological equations, let us consider a flat FLRW metric:
\be
ds^2=-dt^2+a^2(t)\sum_{i=1}^{3} (dx^{i})^2\ ,
\label{FLRWmetric}
\ee
where $a(t)$ is the usual scale factor. As for the energy-momentum tensor for the SM particles and dark matter, we take it to be described in both cases by the one of a perfect fluid, i.e.:
\be
T_{\mu\nu}=(\rho+p)u_{\mu}u_{\nu}+p g_{\mu\nu}\ ,
\ee
where $\rho$ and $p$ account for the energy density and pressure, respectively. The corresponding  FLRW equations for the metric (\ref{FLRWmetric}) with this fluid are easily obtained as (here a dot represents a derivative with respect to time)
\bea
H^2=\left(\frac{\dot{a}}{a}\right)^2=\frac{\kappa^2}{3}\left(\rho_{b}+\rho_{rad}+\rho_{DM}+\rho_{\phi}\right)\ ,\label{FLRWeqs1} \\
\dot{H}=-\frac{\kappa^2}{2}\left(\rho_{b}+\rho_{rad}+\rho_{DM}+p_{rad}+\rho_{\phi}+p_{\phi}\right)\ ,
\label{FLRWeqs2}
\eea
where we have spelled out $\{\rho_b,\rho_{rad},\rho_{DM}\}$ as the energy density for baryons, radiation and dark matter, respectively. Note that we are assuming here a null pressure for both baryons, $p_{b}=0$, and dark matter, $p_{DM}=0$. From (\ref{ContEqs}), the continuity equations for these components yield the result
\be
\dot{\rho}_{rad}+3H (\rho_{rad}+p_{rad})=0\ , \quad \dot{\rho}_{b}+3H \rho_{b}=0\ , \quad \dot{\rho}_{DM}+3H \rho_{DM}=-Q\dot{\phi}\ , 
\label{conteqs2}
\ee
As for the scalar field, its energy density and pressure appearing in Eqs.(\ref{FLRWeqs1}) and (\ref{FLRWeqs2}) are given by
\be
\rho_{\phi}=\frac{1}{2}\dot{\phi}+V(\phi)\ , \quad p_{\phi}=\frac{1}{2}\dot{\phi}-V(\phi)\ ,
\label{phiRhoP}
\ee
respectively, while its equation of motion (\ref{ScalarFeq}) reads, in the FLRW background:
\be
\ddot{\phi}+3H\dot{\phi}+\frac{dV(\phi)}{d\phi}=Q\ .
\label{scalarEqFLRW}
\ee
The set of equations given by (\ref{FLRWeqs1}), (\ref{FLRWeqs2}), (\ref{conteqs2}) and (\ref{scalarEqFLRW}) describe completely the cosmological evolution for a particular form of the scalar potential and the disformal transformation (\ref{disformal}). Actually the scalar field equation (\ref{scalarEqFLRW}) is not independent with respect to the others and can be used instead as an auxiliary equation. Any comoving observer as described by the FLRW metric (\ref{FLRWmetric}) would measure a Hubble expansion rate $H$ governed by the above set of equations, whose proper time coincides with the time coordinate $t$ while their geodesics are expressed in terms of the Christoffel symbols given by the metric (\ref{FLRWmetric}). However, for a dark matter ``observer", i.e., in the frame corresponding to dark matter, the expansion rate as measured by such a hypothetical observer will differ with respect to the other frame. By assuming the following metric in such a frame:
\be
d\tilde{s}^2=\tilde{g}_{\mu\nu}dx^{\mu}dx^{\nu}=-d\tilde{t}^2+\tilde{a}^2(\tilde{t})\sum_{i=1}^{3} (dx^{i})^2\ ,
\label{DisfFLRWmetric}
\ee
and using the disformal transformation (\ref{disformal}), the time coordinates and scale factors in both frames are related as follows:
\be
d\tilde{t}=\sqrt{C-D\dot{\phi}}\ d t\ ,\quad  \tilde{a}(\tilde{t})=\sqrt{C}\ a(t)\ .
\label{tatildes}
\ee
Hence, the expansion rate as measured from each frame might look different, depending on the choice of the disformal transformation (\ref{disformal}). This will become very relevant for some cases analyzed along this paper, as shown below.

\section{Reconstructing cosmological solutions}
\label{Recons}

In this section we shall describe a simple way to reconstruct the corresponding gravitational action (\ref{action}) and the disformal transformation (\ref{disformal}) for a given Hubble solution, to be put to good use later. The scalar field can be redefined in such a way that the kinetic term in the action (\ref{action}) is changed in the following way \cite{Nojiri:2005pu,Elizalde:2008yf}:
\be
\partial_{\mu}\phi\partial^{\mu}\phi \rightarrow \omega(\phi)\partial_{\mu}\phi\partial^{\mu}\phi\ .
\label{scalarbis}
\ee
Under this transformation, the scalar field equation (\ref{scalarEqFLRW}) becomes
\be
\omega(\phi)\ddot{\phi}+3H\omega(\phi)\dot{\phi}+\frac{1}{2} \omega'(\phi)\dot{\phi}^2+V'(\phi)=\tilde{Q}\ ,
\label{scalarEqbis}
\ee
where $\tilde{Q}\equiv \sqrt{\omega}Q$. In this way, we can assume a particular ansatz for the scalar field $\phi$ and, consequently, the dynamics of the field will be encoded in the kinetic term $\omega(\phi)$. The interacting $Q$-term in Eq.(\ref{Qterm}) is now given by
\be
Q=\frac{T^{\mu\nu}_{DM}}{C(1+\frac{D}{C}\omega\partial_{\sigma}\phi\partial^{\sigma}\phi)}\left[\frac{1}{2\sqrt{\omega}}C'g_{\mu\nu}+\left(-\frac{\sqrt{\omega}}{2}D'+\sqrt{\omega}\frac{DC'}{C}-\frac{\omega'}{2\sqrt{\omega}}D\right)\partial_{\mu}\phi\partial_{\nu}\phi-\sqrt{\omega}D\nabla_{\mu}\nabla_{\nu}\phi\right]\ .
\label{Qbis}
\ee
Note that from the two first continuity equations in (\ref{conteqs2}) one easily obtains
\be
\rho_{b}=\rho_{b0}a^{-3}\ , \quad \rho_{r}=\rho_{r0}a^{-4}\ ,
\label{bradsol}
\ee
for baryons and radiation, respectively, with $\rho_{b0}$ and $\rho_{r0}$ its energy densities at the present time. Similarly,  the continuity equation for dark matter in (\ref{conteqs2}) yields:
\be
\dot{\rho}_{DM}+3H \rho_{DM}=-\tilde{Q}\dot{\phi}\ .
\label{DMeqbis1}
\ee
As for the disformal transformation (\ref{disformal})  it now reads as
\be
\tilde{g}_{\mu\nu}=C(\phi)g_{\mu\nu}+\omega(\phi)D(\phi)\partial_{\mu}\phi\partial_{\nu}\phi=C(\phi)g_{\mu\nu}+\tilde{D}(\phi)\partial_{\mu}\phi\partial_{\nu}\phi\ ,
\label{disformalbis}
\ee
where the disformal function is redefined as $\tilde{D}(\phi)=\omega(\phi)D(\phi)$, such that the interacting term $Q$ given in (\ref{Qbis}) is rewritten in the following way:
\be
\tilde{Q} \equiv \sqrt{\omega}Q=\frac{T^{\mu\nu}_{DM}}{C(1+\frac{\tilde{D}}{C}\partial_{\sigma}\phi\partial^{\sigma}\phi)}\left[\frac{1}{2}C'g_{\mu\nu}+\left(-\frac{\tilde{D}'}{2}+\frac{\tilde{D}C'}{C}\right)\partial_{\mu}\phi\partial_{\nu}\phi-\tilde{D}\nabla_{\mu}\nabla_{\nu}\phi\right]\ ,
\label{Qbis2}
\ee
thus effectively removing the factor $\omega$ from it.

From now on we shall omit the tildes over $D$ and $Q$ for sake of notation. Similarly as in the previous section, the set of equations given by the FLRW equations (\ref{FLRWeqs1}) and (\ref{FLRWeqs2}), together with the continuity equations (\ref{conteqs2}) and (\ref{DMeqbis1}) form an independent set, while the scalar field equation (\ref{scalarEqbis}) can be obtained by combinations of the set itself. The advantage obtained from the redefinition of the scalar field (\ref{scalarbis}) is that one can always fix the scalar field as $\phi=t$, so its dynamics is then encoded in the kinetic term $\omega(\phi)$. By doing so, the corresponding solution for a given scalar potential and the conformal/disformal transformation are easily obtained. The inverse process is also useful for reconstructing the theoretical model for a given solution $H(t)$, as shown in some previous works \cite{Elizalde:2008yf}. This way, given the solution:
\be
H=f(t)\ , \quad \rho_{DM}=g(t)\ , \quad \phi=t\ .
\label{sol}
 \ee
where by consistence $a(t)=a_0\e^{F(t)}$ with $F'(t)=f(t)$, and we can choose $a_0=\e^{-F(t_0)}$ to keep $a(t_0)=1$, the following scalar potential and kinetic term are obtained
\bea
V(\phi)&=&\frac{1}{\kappa^2}\left[3f^2(\phi)+f'(\phi)\right]-\frac{1}{2}\left[g(\phi)+\rho_{b0}\left(a_0\e^{F(\phi)}\right)^{-3}+\rho_{r0}\left(a_0\e^{F(\phi)}\right)^{-4}\right] \label{ReconsWV1} \ ,\\
\omega(\phi)&=&-\frac{2}{\kappa^2}f'(\phi)-\left[g(\phi)+\rho_{b0}\left(a_0\e^{F(\phi)}\right)^{-3}+\rho_{r0}\left(a_0\e^{F(\phi)}\right)^{-4}\right] \ .
\label{ReconsWV2}
\eea
Let us point out that the conformal and disformal functions $C(\phi)$ and $D(\phi)$ are given by means of the continuity equation for dark matter (\ref{DMeqbis1}), and they will be obviously degenerated. In the next section we shall proceed to reconstruct some particular solutions containing some usual cosmological singularities, and the corresponding counterpart in the dark matter frame will be obtained and analyzed.

\section{Singular versus regular expansions}
\label{singu}

The occurrence of singularities in the fabric of space-time seems inherent to gravitation, arising from black holes to cosmology. The most widely accepted notion to capture their meaning is that of \emph{geodesic completeness}, namely, whether any time-like and null geodesic can be extended to arbitrarily large values of their corresponding affine parameter (to the future and to the past). This is so because the trajectories of idealized physical observers in the absence of non-gravitational interactions are identified with time-like geodesics. Such observers should be able to observer and interact with the universe at all times and, therefore, should be eternal, i.e., they should not come into existence or be destroyed at any specific finite times. Something similar should happen with null geodesics, which represent the propagation of information. Since information, classical or quantum, should never be destroyed, null geodesics should be defined for all values of their affine parameter. The incompleteness of null geodesics signals the possibility of creating or destroying information, while the incompleteness of time-like geodesics indicates the impossibility of performing physical measurements.

The above notion is ingrained in some theorems formulated in the seventies to provide a clear mathematical formulation of the conditions under which a space-time will hold geodesically incomplete curves \cite{Theorems}. For the sake of this paper we shall focus on the analysis of the occurrence of two types of well known singularities in cosmology  (for a complete list of cosmological singularities see Refs.\cite{Nojiri:2005sx,TV,BeltranJimenez:2016dfc}), in order to determine how they are seen by the geodesic observers in the ordinary and dark matter frames, respectively. Such singularities corresponds to  
\begin{itemize}
  \item ``Big Bang" singularity \cite{BigBang}: For $t \rightarrow t_0$, $a \rightarrow 0$, $\rho \rightarrow \infty$ and $\vert p \vert \rightarrow \infty$.
  \item ``Big Rip" singularity \cite{BigRip}: For $t \rightarrow t_s$, $a \rightarrow \infty$, $\rho \rightarrow \infty$ and $ \vert p \vert  \rightarrow \infty$.
\end{itemize}
Such two singularities are chosen on the grounds of corresponding to geodesically incomplete space-times. This can be proven by studying the geodesic equation
\be \label{geodesicsEQ}
\frac{\dd^2 x^\mu}{\dd \lambda^2}+\Gamma^\mu_{\alpha\beta}\frac{\dd x^\alpha}{\dd \lambda}\frac{\dd x^\beta}{\dd \lambda}=0 \ ,
\ee
where the curve $\gamma^{\mu}=x^{\mu}(\lambda)$ depends on the affine parameter $\lambda$ and $\Gamma^\mu_{\alpha\beta}$ are the Christoffel symbols. For a FLRW spacetime, this yields two sets of equations
\bea
\frac{\dd^2 t}{\dd \lambda^2}+Ha^2\delta_{ij}\frac{\dd x^i}{\dd \lambda}\frac{\dd x^j}{\dd \lambda}&=&0\, ,\label{timegeo}\\
\frac{\dd^2 x^i}{\dd \lambda^2}+2H\frac{\dd x^i}{\dd \lambda}\frac{\dd t}{\dd \lambda}&=&0 \quad \rightarrow \frac{\dd}{\dd \lambda}\left(a^2\frac{\dd x^i}{\dd \lambda}\right)=0\label{spacegeo} \ .
\eea
The second set of equations  (\ref{spacegeo}) can be easily integrated to give:
\be
\frac{\dd x^i}{\dd \lambda}=\frac{k^i_0}{a^2(t)} \ ,
\ee
with $k^i_0$ integration constants. This way, the geodesic equation for the time coordinate (\ref{timegeo}) can be written as:
\be
\left(\frac{\dd t}{\dd\lambda}\right)^2=\frac{\vert\vec{k}_0\vert^2}{a^2(t)}+C_0 \ ,
\label{tlambdageodesic}
\ee
where $C_0$ is another integration constant. As far as the scale factor does not diverge or vanishes for a finite time coordinate, the tangent vector has a regular behaviour and geodesics are complete. This is not the case for the two types of singularities described above. For the Big Bang singularity, the universe meets an incomplete geodesic at a finite past time in which the scale factor goes to zero,  while the Big Rip singularity produces the opposite effect, namely, the scale factor becomes infinite at a finite future time where geodesics meet their end. In geometric scenarios with two metrics in which the standard matter is coupled to one of the frames and the dark matter to the other, like in our case, the discussion of geodesic completeness in the two frames becomes particularly relevant, as it could happen that not both of them need be singular simultaneously. Avoiding incompleteness in one frame may still allow for a consistent combined physical description despite the apparent limitations of a single frame theory. Thus, besides the interesting technical characteristics of this kind of models for model-building,  they may also bring interesting conceptual advantages for the physical interpretation of the theory.  To analyze this question within the framework of the interacting models  described by the action (\ref{action}), for the sake of simplicity we shall neglect radiation and assume a purely conformal transformation between the two frames, i.e., $D(\phi)=0$ in (\ref{disformal}), though its extension to fully disformal transformations is straightforward. In this case, the interacting term (\ref{Qbis2}) reads
\be
Q=\frac{C'}{2C}g^{\mu\nu}{T}_{\mu\nu}^{DM}=-\frac{C'}{2C}{\rho}_{DM}\ .
\label{QConf}
\ee
Let us consider the following solution for the Hubble parameter:
\be
H(t)=\frac{2}{3t}\quad \rightarrow\quad a(t)\propto t^{2/3}\ .
\label{model1}
\ee
This solution is well known to describe a pressureless matter-dominated universe with an initial singularity located for convenience at $t=0$. By assuming the following ansatz on the scalar field and dark matter energy density:
\be
\phi=t\ , \quad \rho_{DM}=\rho_{DM0}\left(\frac{t_0}{t}\right)^3\propto a^{-9/2}\ ,
\ee
the corresponding kinetic term and scalar potential are obtained from (\ref{ReconsWV1}) and  (\ref{ReconsWV2}) as
\bea
\omega(\phi)&=&\frac{4-3\kappa^2t_0^2\rho_{0m}}{3\kappa^2}\frac{1}{\phi^2}-\rho_{DM0}\left(\frac{t_0}{\phi}\right)^3\ , \nn
V(\phi)&=&\frac{1}{2}\omega(\phi)=\frac{4-3\kappa^2t_0^2\rho_{0m}}{6\kappa^2}\frac{1}{\phi^2}-\frac{\rho_{DM0}}{2}\left(\frac{t_0}{\phi}\right)^3\ .
\label{KinetVmodel1}
\eea
As can be easily seen, this leads to a pressureless scalar field, $p_{\phi}=0$, nothing surprising in our universe dominated by a pressureless fluid. Then, the corresponding conformal transformation $C(\phi)$ is obtained through the continuity equation for dark matter (\ref{DMeqbis1}) as
\be
C(\phi)=\frac{1}{9\tilde{H}_0^2}\frac{1}{\phi^2}\ ,
\label{Cmodel1}
\ee
where $\tilde{H}_0$ is an integration constant intentionally written in such a way. Using this conformal function, the corresponding time coordinate and scale factor as measured by an observer at rest in the frame of dark matter (\ref{tatildes}) are obtained as
\be
\tilde{t}=\tilde{t}_0+\frac{1}{3\tilde{H}_0}\log t\ , \quad \tilde{a}(\tilde{t})=\tilde{a}_0\e^{-\tilde{H}_0(\tilde{t}-\tilde{t}_0)}\ ,
\label{tatildesmodel1}
\ee
where $\tilde{t}_0$ is an integration constant. Then, the Hubble rate as measured from such an observer remains constant (and negative), i.e.:
\be
\tilde{H}(\tilde{t})=-\tilde{H}_0\ .
\label{Htildemodel1}
\ee
Hence, the time in which the Big Bang singularity occurs in the ordinary matter frame, $t=0$, corresponds in the dark matter frame to $\tilde{t}\rightarrow-\infty$ and, moreover, the scalar factor remains finite and does not vanish in such a frame,  while its geodesics (\ref{tlambdageodesic}) remain extensible to arbitrarily large times. Therefore, what for an observer in the SM frame is a singularity (i.e. a geodesically incomplete universe),  for an observer in the dark matter frame the universe is eternal and contracting.

A second example correspond to the future singularity of the Big Rip. Let us assume the following ansatz for the Hubble rate in this case:
\be
H(t)=\frac{1}{t_s-t}\quad \rightarrow\quad a(t)\propto \frac{1}{t_s-t}\ .
\label{model2}
\ee
This solution describes a ``phantom''-dominated universe with a Big Rip occurring at $t=t_s$. As in the previous case, we assume a particular solution for the scalar field and the dark matter fluid given by:
\be
\phi=t\ , \quad \rho_{DM}=\rho_{DM0}\left(\frac{t_s-t}{t_s-t_0}\right)^2\propto a^{-2}\ .
\label{phiDM2}
\ee
Here $t_0$ is chosen such the dark matter density measured at $t=t_0$ is given by $\rho_{DM0}$. Hence, by the expressions (\ref{ReconsWV1}) and (\ref{ReconsWV2}) the kinetic term and potential for the scalar field are obtained as
\bea
\omega(\phi)&=&\frac{-4}{\kappa^2}\frac{1}{(t_s-\phi)^2}-\left[\rho_{m0}\left(\frac{t_s-\phi}{t_s-t_0}\right)^3+\rho_{DM0}\left(\frac{t_s-\phi}{t_s-t_0}\right)^2\right]\ , \nn
V(\phi)&=&\frac{-2}{\kappa^2}\frac{1}{(t_s-\phi)^2}-\frac{1}{2}\left[\rho_{m0}\left(\frac{t_s-\phi}{t_s-t_0}\right)^3+\rho_{DM0}\left(\frac{t_s-\phi}{t_s-t_0}\right)^2\right]\ ,
\label{KinetVmodel1}
\eea
respectively, while the interacting term yields
\be
Q=-\frac{1}{t_s-\phi}\rho_{DM}\ ,
\ee
and, consequently, the conformal transformation is given by
\be
C(\phi)=\frac{4}{\tilde{H}_0^2}\frac{1}{(t_s-\phi)}\ ,
\label{Cmodel2}
\ee
where $\tilde{H}_0$ is an integration constant written in this way as in the previous case since it corresponds to the value of the Hubble parameter as measured from the dark matter frame. Hence, the corresponding time and scale factor as measured in the dark matter frame (\ref{tatildes}) are
\be
\tilde{t}=\tilde{t}_0-\frac{2}{\tilde{H}_0}\log (t_s-t)\ , \quad \tilde{a}(\tilde{t})=\tilde{a}_0\e^{\tilde{H}_0(\tilde{t}-\tilde{t}_0)}\ .
\label{tatildesmodel2}
\ee
Finally the Hubble parameter in the dark matter frame yields
\be
\tilde{H}(\tilde{t})=\tilde{H}_0\ .
\label{Htildemodel2}
\ee
As seen by inspection of (\ref{tatildesmodel2}), the Rip time in the ordinary frame, $t=t_s$, corresponds to $\tilde{t}\rightarrow\infty$ in the dark matter frame. As in the previous case, an observer in such a frame does not notice any singularity in its frame but an universe that expands exponentially as described by the Hubble parameter (\ref{Htildemodel2}). Actually its proper time covers the full infinite range of its comoving time, $-\infty<\tilde{t}<\infty$, such that the universe as seen from such frame is eternal and the geodesics are completely regular. 

The bottom line of the discussion on the two examples above is that  for an appropriate coupling between dark energy and dark matter, described by the conformal function $C(\phi)$, the universe expansion will be singular or regular depending on the frame each observer is living at. This follows simply from the fact that by the assumption (\ref{action}), different species follow geodesics described by different metrics, such that the geometry as seen from one or another frame turns out to be different to such observers. While these results have been obtained for arguably toy models constructed ad hoc just for showing this theoretical possibility, in the next section we shall assume some particular phenomenological parametrizations of the Hubble parameter in order to fit them with different sources of data, to discuss whether such a construction might be a good description of the observable universe.

\section{Some phenomenological models}
\label{Phenom}

Let us now consider a phenomenological parametrization of the Hubble parameter that contains the core ideas discussed above on the interacting dark energy-dark matter model described by the action (\ref{action}) in order to be compared with different observational data to check whether the theoretical possibility explored in the previous section is viable. Firstly, we can express the continuity equations for dark matter and the scalar field in Eqs.(\ref{conteqs2}) and (\ref{DMeqbis1}) in terms of the redshift instead of the time, i.e.:
\bea
- (1+z)\rho_{DM}'+3\rho_{DM}=-(1+z) \rho_{DM} \frac{C_{\phi}}{2C}\phi' \label{ContZ1} \\
- (1+z)\rho_{\phi}'+3(\rho_{\phi}+p_{\phi})=(1+z)\rho_{DM}\frac{C_{\phi}}{2C}\phi' \ ,
\label{ContZ2}
\eea
where primes denote derivatives with respect to the redshift $z$. Paralleling our analysis above, we can define the scalar field as $\phi=z$ such that its dynamics is encoded in the kinetic term function $\omega(z)$. Unless $C(\phi)=$constant, that recovers the uncoupled case, the dark matter energy density will deviate from the usual $\rho_{DM}\propto (1+z)^3$ of the $\Lambda$CDM model. Hence, we can assume that the dependence on the redshift goes as a different power of the redshift:
\be
\rho_{DM}=\rho_{DM0}(1+z)^{3+m/2}\ ,
\label{DMZ}
\ee
where $m$ is a free parameter that accounts for deviations with respect to the CDM model and is the target to be fit with the observational data. For simplicity, we shall also assume that dark energy (i.e., the scalar field energy density) follows another power law of the redshift:
\be
\rho_{\phi}=\rho_{\phi0}(1+z)^{n}\ ,
\label{phiZ}
\ee
being $n$ another free parameter. Hence, by the first Friedmann equation (\ref{FLRWeqs1}), the corresponding Hubble parameter is obtained as
\be
H^2(z)=H_0^2\left[\Omega_{DM}(1+z)^{3+m/2}+\Omega_{\phi}(1+z)^{n}\right]\ ,
\label{HZ1}
\ee
where $\Omega_{i}=\frac{\rho_{i0}}{3H_0^2/\kappa^2}$ are the usual cosmological parameters and, for coherence, $1=\Omega_{DM}+\Omega_{\phi}$. 

For the phenomenological model above we have neglected the radiation and baryon contributions, so that we have as free parameters $\{n,m,\Omega_{DM}\}$. Nevertheless, we may impose some additional restrictions - priors - on $\{n,m\}$ in order to avoid large correlations that naturally arise as can be easily seen by inspection of (\ref{HZ1}). Hence, we are considering as a first testable model the Hubble parameter given in (\ref{HZ1}) and assuming a gaussian prior on the parameter $m$ centered in zero. This is a natural choice as one expects small deviations from the $\Lambda$CDM model. On the other hand, note that as far as $n<0$ a Big Rip singularity occurs (in the frame of the standard matter fields). Whether or not this singularity is noticed by an observer in the dark matter frame will depend on the conformal coupling $C(\phi)$ which can be easily obtained from (\ref{ContZ2}) and (\ref{DMZ}), leading to:
\be
C(\phi)=C_0(1+\phi)^{m}\ .
\label{CZ}
\ee
For an observer in the dark matter frame, the redshift $\tilde{z}$ measured for a photon propagating towards the observer is related to the redshift measured by the ordinary observer as:
\be
1+ \tilde{z}=\frac{\tilde{a}_0}{\tilde{a}(\tilde{t})}=\frac{\tilde{a}_0}{a_0}\frac{1}{\sqrt{C(z)}}(1+z)\ ,
\label{Zzz}
\ee
where $\tilde{a}_0$ is the scale factor measured at $\tilde{t}_0$ by the dark matter observer while $a_0$ is the one measured by the ordinary observer at $t_0$. The Hubble parameters for both frames are related as follows:
\be
\tilde{H}(\tilde{z})=\frac{H(z)}{\sqrt{C(z)}}-(1+z)H(z)\frac{C'(z)}{2\sqrt{C^3(z)}}\ .
\label{HzHZ}
\ee
By using the conformal coupling (\ref{CZ}), the corresponding relation for both redshifts results in
\be
1+\tilde{z}=\zeta(1+z)^{\frac{2}{2-m}}\ ,
\ee
where $\zeta=\left(\sqrt{C_0}\frac{a_0}{\tilde{a}_0}\right)^{\frac{2}{2-m}}$ is a constant. Hence, for the model (\ref{HZ1}) the corresponding Hubble parameter in the dark matter frame can be obtained easily through (\ref{HzHZ}):
\be
\tilde{H}(\tilde{z})=\frac{H_0(2-m)}{2\sqrt{C_0}}\sqrt{\Omega_{DM}\zeta^{\frac{6-m}{2}}(1+\tilde{z})^{\frac{6-m}{2-m}}+\Omega_{\phi}\zeta^{n-m}(1+\tilde{z})^{2\left(\frac{n-m}{2-m}\right)}}\ .
\label{HZzz1}
\ee
For redshifts close to $z\sim0$ (or even negative $-1<z<0$ if one considers $z$ just as the independent variable) and $n<0$ (Big Rip singularity), the term dominating in (\ref{HZ1}) will be the dark energy one:
\be
H(z)\sim H_0\sqrt{\Omega_{\phi}(1+z)^n}\ , 
\label{HOMz}
\ee
while in the dark matter frame the corresponding Hubble parameter reads
\be
\tilde{H}(z)\sim \frac{H_0(2-m)}{2\sqrt{C_0}}\sqrt{\Omega_{\phi}\zeta^{n-m}(1+\tilde{z})^{2\left(\frac{n-m}{2-m}\right)}}\ .
\label{HDMz}
\ee	
For $\frac{n-m}{2-m}\geq0$, there is no future singularity in the dark matter frame even when $n<0$. Hence, as a second testable model we might consider $n=m$, which reduces the number of free parameters and opens up the possibility to have a singular universe in the ordinary matter frame and an eternal one from the point of view of the dark matter frame, where the universe would tend asymptotically to de Sitter. Therefore, the second model considered here is given by:
\be
H^2(z)=H_0^2\left[\Omega_{DM}(1+z)^{3+m/2}+(1-\Omega_{DM})(1+z)^{m}\right]\ .
\label{HZ2}
\ee
This model is just a simplification with respect to the previous model as it involves just two free parameters $\{m,\Omega_{DM}\}$ and it provides a de Sitter expansion in the dark matter frame by construction.

\section{Fittings and results}
\label{results}

In order to fit the free parameters of both models (\ref{HZ1}) and (\ref{HZ2}) with observational data, we use the Union 2.1 SN catalogue  (see \cite{Suzuki:2011hu}) with the data of $N_{\text{SN}}=557$ supernovae of Type Ia. This dataset provides the redshift and the distance modulus for each SNe Ia and the corresponding errors $\sigma_{obs}$. The theoretical distance modulus is given by: 
\begin{equation}
\mu_{\text{theo}}(z;\Omega_{DM},n,m)=\bar\mu+5\log_{10} \left[D_L(z;\Omega_{DM},n,m)\right]\ .
\label{modulus} 
\end{equation}
Here $\bar \mu=-5\log_{10}\left[{H_0\over c}\right]+25$ is a nuisance parameter and $D_L$ is the luminosity distance defined as:
\begin{equation}
D_L(z;\Omega_m,\Omega_{DM},n,m)= (1+z) \int_0^z {\rm d}z'\frac{H_0}{H(z';\Omega_{DM},n,m)}\ .
\label{SN1} 
\end{equation}
We are using Monte Carlo Markov Chains with the Metropolis-Hastings algorithm to explore the free parameters space, which consists on random checks on the parameter space subjected to maximize the likelihood, the latter defined as
\begin{equation}
\mathcal{L}= {\cal N} {\rm e}^{- \chi^2/2}\ . 
\label{SN3} 
\end{equation} 
Here ${\cal N}$ is a normalisation constant and $\chi^2$ is given by
\begin{eqnarray}
\chi^2_{SNe} =\sum_{i=1}^{N} \frac{(\mu_{\text{obs}}(z_i) - \mu_{\text{theo}}(z_i;{\bar \mu}, \Omega_{DM},n,m))^2} {\sigma_{\text{obs}}^2(z_i)} \ .
\label{chi2} 
\end{eqnarray}
The nuisance parameter $\bar \mu$ can be easily marginalised by expanding (\ref{chi2}) and minimizing the expression with respect to  $\bar \mu$ (for more details see Refs.~\cite{Leanizbarrutia:2014xta,Lazkoz:2005sp}). 

In addition, we also use datasets from Baryonic Acoustic Oscillations (BAO \cite{H_BAO}), which provide one of the following two magnitudes:
\begin{equation}
d_z(z)= \frac{r_s(z_d)}{D_V(z)}\, ,\quad
A(z) = \frac{H_0\sqrt{\Omega_{DM}}}{cz}D_V(z)\ .
\label{dzAz}
\end{equation}
Here  $r_s(z_d)$ is the comoving sound horizon at decoupling $z_d$ and is given by:
\be r_s(z)=  \int_z^{\infty}
\frac{c_s(z')}{H (z')}\,dz'=\frac1{\sqrt{3}}\int_0^{1/(1+z)}\frac{da}
 {a^2H(a)\sqrt{1+\big[3\Omega_b^0/(4\Omega_\gamma^0)\big]a}}\ ,
  \label{rs2}
\ee
whereas $D_V(z)$ is:
\be
D_V(z)=\left[\frac{cz D_M^2(z)}{H(z)} \right]^{1/3}\, ,\quad \text{where} \quad
D_M(z)=\frac{D_L(z)}{1+z}= c \int_0^z \frac{d\tilde z}{H (\tilde z)}\, .
\ee
In total we use 17 BAO data points. The $\chi^2$ for the BAO data (\ref{dzAz}) is obtained as follows:
\begin{equation}
\chi^2_{\mathrm{BAO}}(\Omega_{DM},\dots)=\Delta d\cdot C_d^{-1}(\Delta d)^T
+ \Delta { A}\cdot C_A^{-1}(\Delta { A})^T\, ,
\label{chiB}
\end{equation}
where $C_{d}$ and $C_{A}$ are the covariance matrices for the correlated data, while $\Delta d$ and $\Delta A$ are the differences among the observational and the theoretical values. 

The data concerning the Hubble parameter $H(z)$ that is used here corresponds to the set obtained by estimating the age of galaxies at different redshifts \cite{HzData}. The $\chi^2$ is given by:
\begin{equation}
    \chi_H^2(\Omega_{DM},\dots)=\sum_{j=1}^{N_H}\bigg[\frac{H(z_j,\Omega_{DM},\dots)-H^{obs}(z_j)}{\sigma _j}  \bigg]^2
    \label{chiH}
\end{equation}
where we marginalize over the nuisance parameter $H_0$. Then, the total $\chi^2_{Tot}$ is calculated by the sum of the whole set of fittings:
\be
\chi^2_{Tot}=\chi^2_{SNe}+\chi^2_{\mathrm{BAO}}+ \chi_H^2\ .
\label{totalchi2}
\ee

\begin{table}[t!]
\begin{tabular}{|c|c|}

\hline
                     & SNe Ia + BAO + H(z)      \\ \hline
$\Omega_m$            & 0.30$^{+0.009}_{-0.009}$   \\ \hline
$\chi^2_{\text{min}}$ & 580.35\\ \hline
$\chi^2_{\text{red}}$ & 0.96\\ \hline
\end{tabular}
\caption{Mean values for $\Lambda$CDM model with the corresponding errors for the fitting with SNe Ia+BAO+H(z).}
\label{TableLCDM}
\end{table}

\begin{table}[t!]
\begin{tabular}{|c|c|c|c|}
\hline
                      & SNe Ia                  & BAO                      & SNe Ia + BAO + H(z)      \\ \hline
$\Omega_m$            & 0.48$^{+0.23}_{-0.23}$  & 0.28$^{+0.016}_{-0.018}$ & 0.30$^{+0.016}_{-0.016}$ \\ \hline
$m$                   & -1.04$^{+1.18}_{-1.13}$ & -0.056$^{+0.11}_{-0.11}$ & 0.020$^{+0.080}_{-0.086}$ \\ \hline
$\chi^2_{\text{min}}$ &  542.67                   &   9.98                   &     580.23                 \\ \hline
$\chi^2_{\text{red}}$ & 0.98                    & 0.67                     & 0.96                     \\ \hline
\end{tabular}
\caption{Mean values for the free parameters including the corresponding errors for the two parameters model (\ref{HZ2}) for the fittings with SNe Ia, BAO, and SNe Ia+BAO+H(z).}
\label{TableModel1}
\end{table}

\begin{table}[t!]
\begin{tabular}{|c|c|c|c|}
\hline
                      & SNe Ia                   & BAO                      & SNe Ia + BAO + H(z)      \\ \hline
$\Omega_m$            & 0.30$^{+0.09}_{-0.08}$   & 0.28$^{+0.015}_{-0.017}$ & 0.30$^{+0.016}_{-0.016}$ \\ \hline
$m$                   & -0.051$^{+0.31}_{-0.30}$  & 0.034$^{+0.20}_{-0.20}$  & 0.29$^{+0.16}_{-0.15}$   \\ \hline
$n$                   & $-0.34^{+0.78}_{-0.61}$ & -0.11$^{+0.18}_{-0.20}$  & -0.28$^{+0.16}_{-0.16}$  \\ \hline
$\chi^2_{\text{min}}$ &  542.51                    &    9.81                  &         574.49             \\ \hline
$\chi^2_{\text{red}}$ & 0.98                     & 0.70                     & 0.96                     \\ \hline
\end{tabular}
\caption{Mean values for the free parameters including the corresponding errors for the three parameters model (\ref{HZ1}) for the fittings with SNe Ia, BAO, and SNe Ia+BAO+H(z).}
\label{TableModel2}
\end{table}

\begin{figure}[t!]
\centerline{ \includegraphics[scale=0.45]{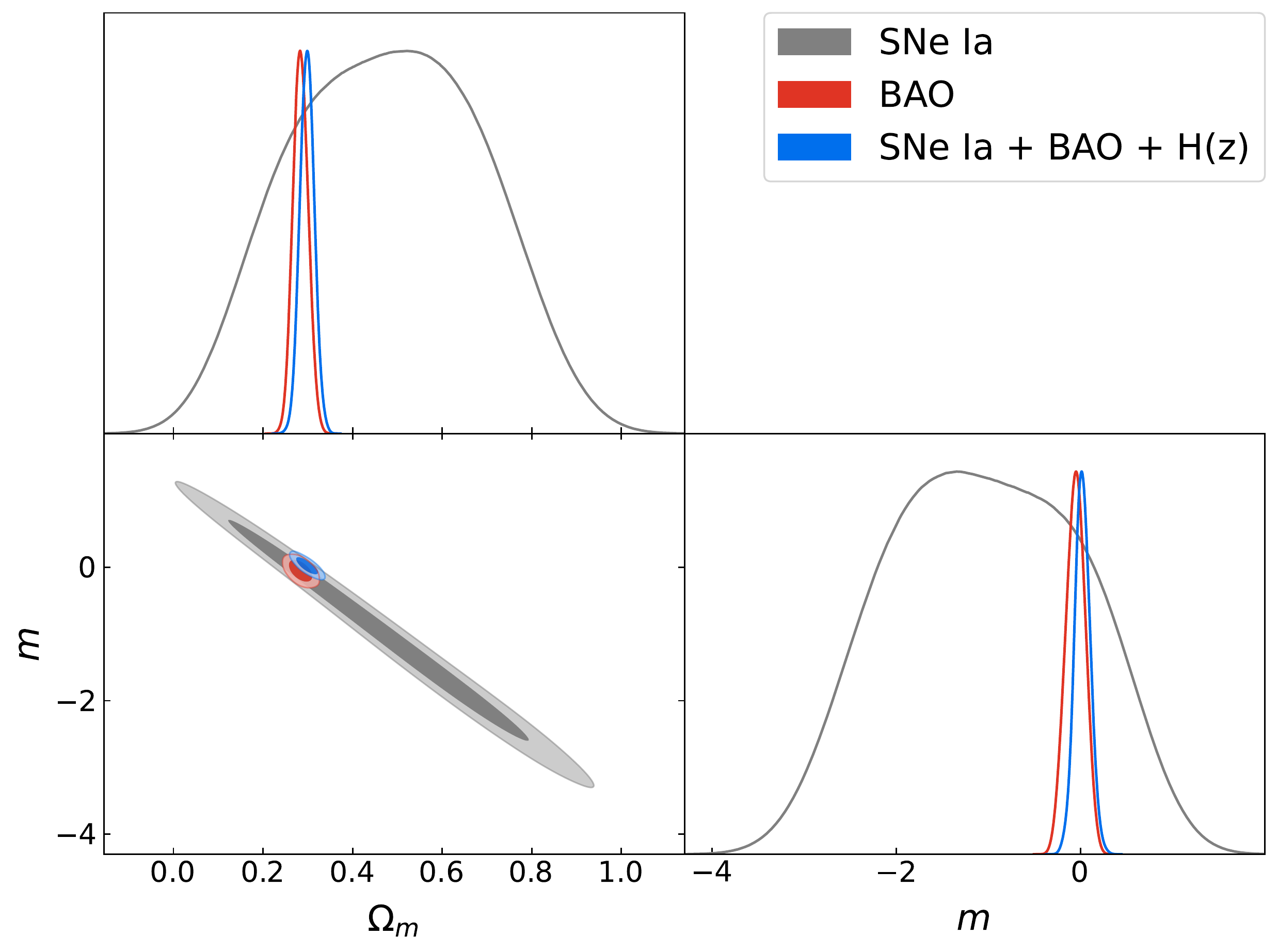}}
\caption{Contour plots and likelihoods for the two parameters model (\ref{HZ2}).} \label{ContourModel2param}
\label{FigModel1}
\end{figure}
\begin{figure}[t!]
\centerline{ \includegraphics[scale=0.45]{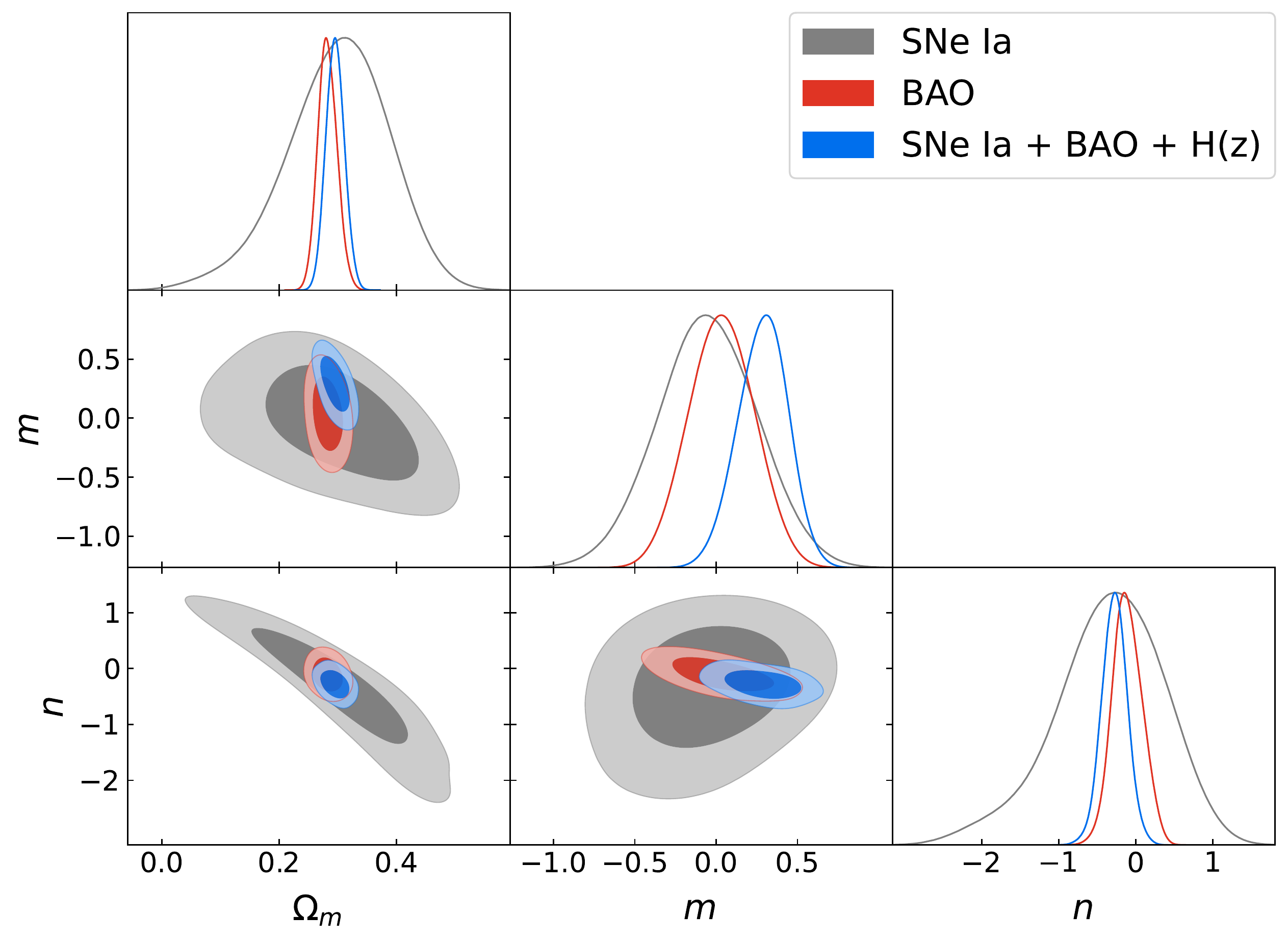}}
\caption{Contour plots and likelihoods for the three parameters model (\ref{HZ1}).} \label{ContourModel3param}
\label{FigModel2}
\end{figure}

The results of the fittings are summarized in Tables \ref{TableModel1} and \ref{TableModel2} for the two-parameter (\ref{HZ2}) and the three-parameter (\ref{HZ1}) models, respectively, where the mean values and the corresponding errors for the parameters are shown together with the $\chi^2_{\text{min}}$ and the reduced $\chi^2$, which is useful for comparing the goodness of fits among different models and is defined as follows:
\be
\chi^2_{\text{red}}=\frac{\chi^2_{\text{min}}}{N-c-1}\ ,
\ee
where $N$ is the number of the data points and $c$ is the number of free parameters of the model. As a comparison, the same fits for the $\Lambda$CDM model are included in Table \ref{TableLCDM}. Moreover, the corresponding contour plots are depicted in Figs.~ \ref{FigModel1} and \ref{FigModel2} for both phenomenological models, where we have used the Python library \textit{getdist} (see Ref.~\cite{Lewis:2019xzd}). As shown in the tables, both models fit the data as good as the $\Lambda$CDM model according to the $\chi^2_{\text{red}}$. Nevertheless, the two parameters model (\ref{HZ2}) exhibits large errors when fit with SNe Ia data for both free parameters, but become better constrained when including BAO and H(z) datasets. It is clear from the expression of this model (\ref{HZ2}) and by inspection of Fig.~ \ref{FigModel1} that both free parameters are correlated and that is the reason behind such large errors. In comparison, the model with three parameters (\ref{HZ1}) shows much smaller errors as shown in Table \ref{TableModel2} and Fig.~\ref{FigModel2}, where we have assumed  a Gaussian prior for the power $m$ in order to avoid even larger correlations than the previous model. Such a prior just assumes that the dependence of dark matter with respect to the redshift does not deviate too much from $\Lambda$CDM, which is perfectly natural, as shown also by the two parameters model (\ref{HZ2}) where such a parameter is let free but fits close to zero when all the  datasets are included in the analysis.

\begin{figure}[t!]
\centerline{
\includegraphics[scale=0.60]{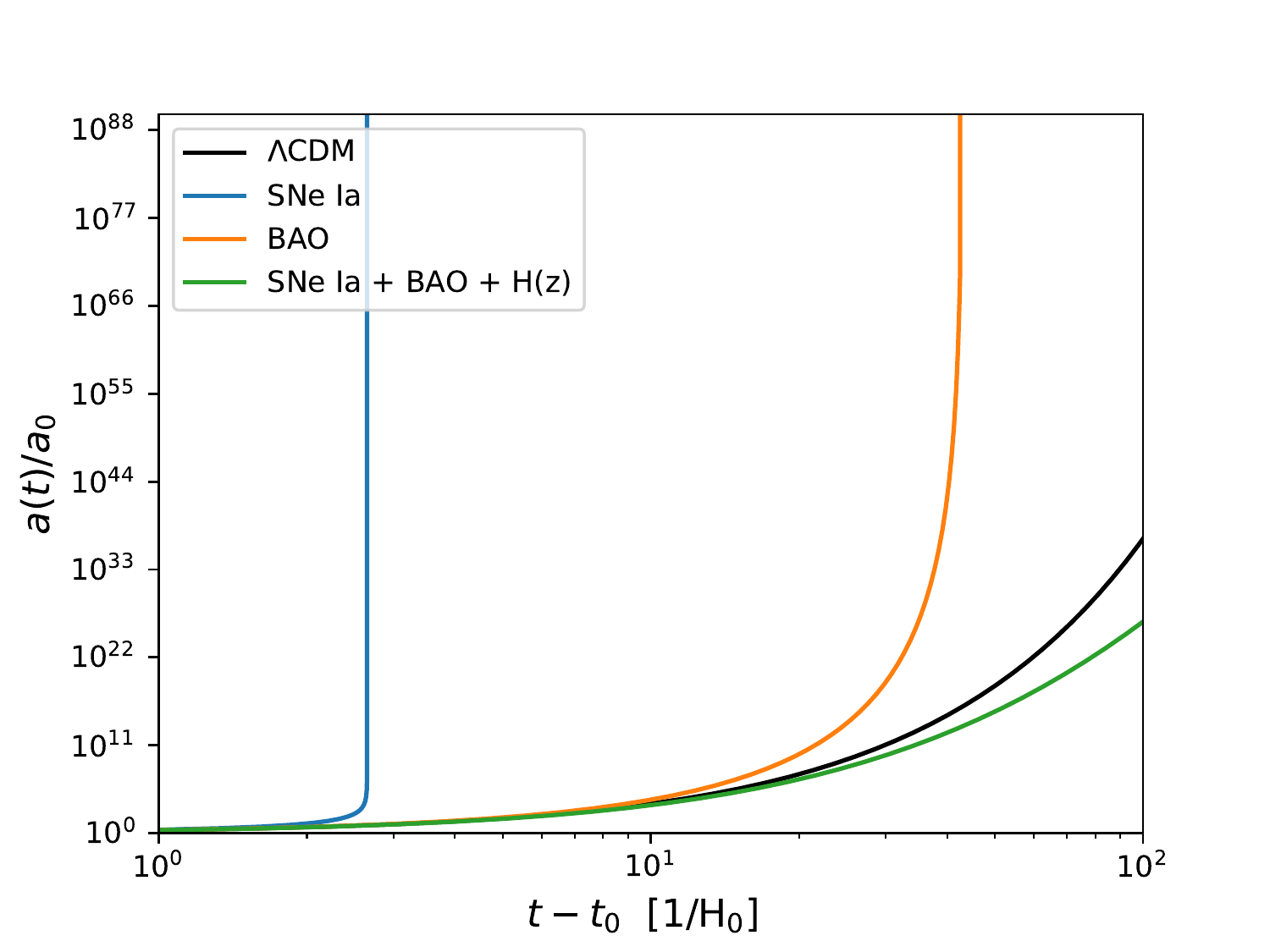}
\includegraphics[scale=0.60]{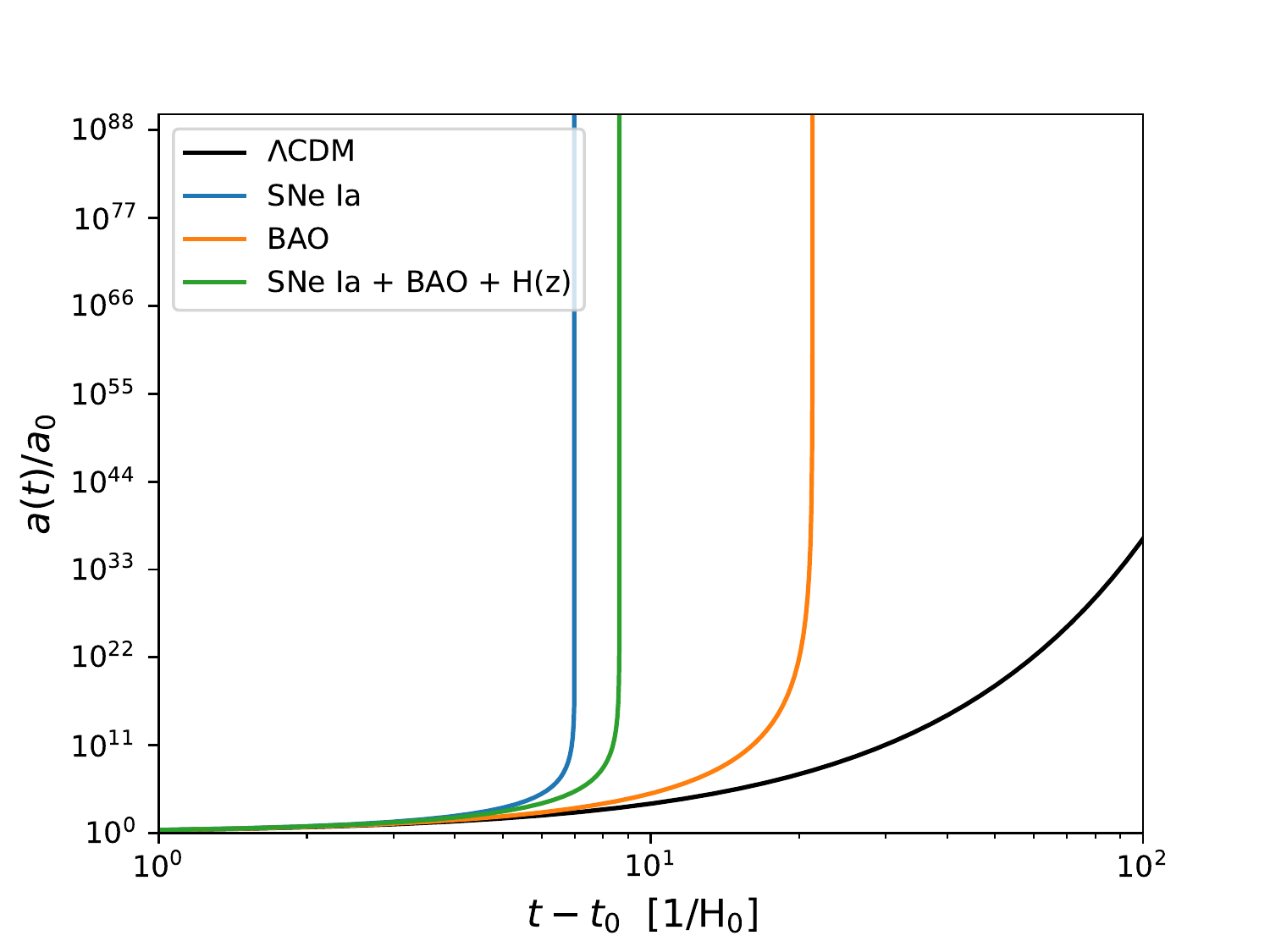}}
\caption{Scale factor evolution for the two parameters model (\ref{HZ2}) (left) and the three parameters model (\ref{HZ1}) (right).}
\label{Figscale}
\end{figure}

As pointed out in the previous section, any negative power for the redshift in the dark energy contribution implies a future Big Rip singularity. In Fig. \ref{Figscale} the future evolution for the scale factor for the two-parameter (\ref{HZ2}) and three-parameter (\ref{HZ1}) models is depicted on the left and right panels, respectively, where we assume the mean values for the free parameters shown in the Tables \ref{TableModel1} and \ref{TableModel2}, respectively. For the two parameters model (\ref{HZ2}), the mean value for $m$ is negative for SNe Ia and BAO, while remains positive when considering all the datasets together, although completely centered in zero. It is clear that one cannot infer the dependence of dark energy by these results (as pointed in many papers before), but for illustrative purposes Fig.~\ref{Figscale} (left panel) shows the Rip time (from $t_0$ and in units of $H_0^{-1}$) for those cases where the singularity occurs, together with the case (SNe Ia+BAO+H(z)) where no singularity occurs. The Rip time is in agreement with other previous analysis where the possibility of a Big Rip singularity is considered \cite{BeltranJimenez:2016fuy}. By construction, this model leads to a regular and constant Hubble parameter in the dark matter frame and the Rip time corresponds to an infinite proper time for dark matter observers, which therefore do not observe any time-like singularity throughout their entire cosmological history. 

The case for the three parameters model (\ref{HZ1}) shows that the corresponding mean for the power of the dark energy contribution is less than zero for all the datasets, which corresponds to a Big Rip singularity. Such a possibility suggested by every dataset is obviously not conclusive statistically. The corresponding future evolution for the scale factor is depicted in Fig.~\ref{Figscale} (right panel), which shows the Rip time for every fit. Whether this singularity is felt from the dark matter frame or not depends on the relative values of $m/n$ as can be easily inferred from the Hubble expression (\ref{HZzz1}), which neglecting the dark matter contribution will go as $(1+\tilde{z})^{\frac{n-m}{2-m}}$, such that for the mean values of Table \ref{TableModel2}, a Big Rip singularity will also occur in the dark matter frame. Nevertheless, as in the previous model, the errors are large enough to include the possibility to have a eternal universe as far as $\frac{n-m}{2-m}\geq0$. 

\section{Conclusions}
\label{conclusions}

In the present paper we have analyzed a novel perspective in the framework of interacting dark energy-dark matter models that are constructed from conformal and disformal couplings between their corresponding metric tensors. Because of this construction, such a type of models induces a fifth force on the dark matter component due to the interaction with dark energy, in such a way that the former does not follow the same space-time geodesics as the latter (and as SM particles), while its dependence on the scale factor (redshift) departs from the usual Cold Dark Matter model. While this type of models have been widely analyzed  in the literature on dark matter model building, in particular being shown to be perfectly compatible with observational data, in this work we focused on conceptual aspects related to the potential development of singularities on each (standard and dark matter) frame within them.

By establishing a standard procedure to obtain analytical solutions for the simpler case of conformal couplings (which can be easily extended to fully disformal ones), we have obtained several solutions holding the two most worrying kinds of cosmological singularities, namely, the  Big Bang and the Big Rip one, as defined by their geodesically incomplete character. We have shown that for the appropriate conformal coupling, both types of singularities that occur in the frame of ordinary SM observers do not necessarily occur in the dark matter frame, since the trajectories for the observers in the latter are perfectly regular and extendible to infinite values of their affine parameter. Hence, while an ordinary SM observer will experience a singularity on its past/future light cone according to its perspective,  the universe will be eternal in the dark matter frame. This might have implications for the understanding of singular space-times in gravitation, as the presence of fifth forces (in this case felt by the hypothetical dark matter observer) can make the job of partially healing a particular space-time, as there might exist families of observers for which information is always accessible (never created nor destroyed) and physical observations always possible.  

In order to investigate the observational viability of the models based on these principles, we have proposed two simple phenomenological implementations of them in which a singularity may arise in one frame but be absent in the other. We have shown that for some subsets of both cases (as defined by their corresponding model parameters) the occurrence of a future Big Rip in the standard matter frame is compatible with observations (SNe Ia, BAO and H(z)), while the expansion remains regular from the point of view of an observer in the dark matter frame. This effectively proves that the latter observers actually see an eternal universe in comparison to the singular one viewed by the ordinary observers. The bottom line of this analysis is that singularities can be removed by a suitable choice of the corresponding interacting term between the dark components while the corresponding cosmological background evolution is still compatible with current data, a result that might help to get a better grip on the actual meaning of cosmological singularities. Moreover, since geodesics and curvature scalars are not invariant under conformal/disformal transformations, the very notion of singularity, under these perspectives, is also not immutable. In this work we have highlighted this fact by working out the dark matter sector in order to alleviate conceptual problems that afflict the standard GR description.

\section*{Acknowledgements}

DAO is funded by the JAE programme of CSIC (Spain). DRG is funded by the {\it Atracci\'on de Talento Investigador} programme of the Comunidad de Madrid (Spain) No. 2018-T1/TIC-10431. DS-CG is funded by the University of Valladolid (Spain), Ref. POSTDOC UVA20.  This work is supported by the Spanish Grants FIS2017-84440-C2-1-P, PID2019-108485GB-I00, PID2020-116567GB-C21 and PID2020-117301GA-I00 funded by MCIN/AEI/10.13039/501100011033 (``ERDF A way of making Europe" and ``PGC Generaci\'on de Conocimiento"), the project PROMETEO/2020/079 (Generalitat Valenciana), the project H2020-MSCA-RISE-2017 Grant FunFiCO- 777740, the project i-COOPB20462 (CSIC),  the FCT projects No. PTDC/FIS-PAR/31938/2017 and PTDC/FIS-OUT/29048/2017, and the Edital 006/2018 PRONEX (FAPESQ-PB/CNPQ, Brazil, Grant 0015/2019). This article is based upon work from COST Action CA18108, supported by COST (European Cooperation in Science and Technology).

\end{document}